\documentclass[aps,floats,prb,twocolumn,showpacs]{revtex4}
\usepackage{amsfonts}
\usepackage{amssymb}
\usepackage{amsmath}
\usepackage[final,dvips]{graphicx}
\begin{document}
\title{Upper critical field in dirty two-band superconductors: breakdown of the
anisotropic Ginzburg-Landau theory}
\author{A.\ A.\ Golubov}
\affiliation{Faculty of Science and Technology, University of Twente, 7500 AE Enschede,
The Netherlands}
\author{A.\ E.\ Koshelev}
\affiliation{Materials Science Division, Argonne National
Laboratory, Argonne, Illinois 60439} \keywords{}
\date{\today }

\begin{abstract}
We investigate the upper critical field in a dirty two-band
superconductor within quasiclassical Usadel equations. The regime
of very high anisotropy in the quasi-2D band, relevant for
MgB$_{2}$, is considered. We show that strong disparities in
pairing interactions and diffusion constant anisotropies for two
bands influence the in-plane $H_{c2}$ in a different way at high
and low temperatures. This causes temperature-dependent $H_{c2}$
anisotropy, in accordance with recent experimental data in
MgB$_{2}$. The three-dimensional band most strongly influences the
in-plane $H_{c2}$ near $T_{c}$, in the Ginzburg-Landau (GL)
region. However, due to a very large difference between the c-axis
coherence lengths in the two bands, the GL theory is applicable
only in an extremely narrow temperature range near $T_c$. The
angular dependence of $H_{c2}$ deviates from a simple
effective-mass law even near $T_c$.

\end{abstract}
\maketitle

\section{Introduction}

There is a strong evidence of the multigap nature of
superconducting state in the recently discovered \cite{Akimitsu}
compound MgB$_{2}$. The concept of multiband superconductivity was
introduced in \cite{Suhl,Moskal} for the case of large disparity
of the electron-phonon interaction for the different Fermi-surface
sheets. For MgB$_{2}$, first-principles calculations of the
electronic structure and the electron-phonon interaction
\cite{Kortus,An,LiuPRL01,Shulga,Kong,Yildirim} have revealed two
distinct groups of bands, namely strongly superconducting
quasi-two-dimensional $\sigma$-bands and weakly superconducting
three-dimensional $\pi$-bands. Quantitative predictions for
various thermodynamic and transport properties of MgB$_{2}$ were
made in the framework of the two-band model.
\cite{Choi,Golub,Brink,Mazin02}

A large number of experimental data, in particular tunneling,
\cite{GiubileoPRL01,IavaronePRL02} point contact
measurements,\cite{SzaboPRL01,SchmidtPRL01,Gonnelli} and heat
capacity measurements,\cite{BouquetPRL01} directly support the
concept of a double gap MgB$_{2}$. Intraband impurity scattering
in both bands may vary in large limits, while interband scattering
is always weak due to the disparity of $\sigma$- and $\pi$-band
wave functions.\cite{Mazin02} This explains the extremely weak
suppression of $T_{c}$ by impurities and the weak correlation
between $T_{c}$ and the resistivity. Therefore, a unique feature
of the MgB$_{2}$ is that the two-gap nature of superconductivity
persists even in the dirty limit for the intraband scattering
rates.

Superconductivity in the two bands is characterized by different
energy and length scales which show up in several properties of a
superconductor. Particularly interesting are the properties of the
mixed state. The c-axis Abrikosov vortex structure in MgB$_{2}$
was studied by STM in Ref.\ \onlinecite{Eskildsen02}, which probes
mainly the weakly superconducting $\pi$-band.  A large vortex core
size compared to estimates based on $H_{c2}$ and the rapid
suppression of the apparent tunneling gap by small magnetic fields
has been reported. These observations can be naturally explained
within the two-band model.\cite{Nakai,KG}

One of the most spectacular consequences of the two-band
superconductivity is the unusual behavior of anisotropy factors
for different physical parameters.\cite{KoganBudko} It was
demonstrated that in clean MgB$_2$ samples the anisotropy of the
London penetration depth,\cite{Kogan_lam,Golub_lam}
$\gamma_{\lambda}$, has to be very different from the anisotropy
of the upper critical field,\cite{Kogan,Dahm} $\gamma_{c2}$. Both
anisotropy factors should strongly depend on temperature and have
opposite temperature dependencies: $\gamma_{\lambda}$ is expected
to increase and $\gamma_{c2}$ is expected to decrease with
temperature. Strong temperature dependence of $\gamma_{c2}$ has
been reliably confirmed by
experiment.\cite{Sologub,Budko,Angst,Eltsev,Lyard,WelpPRB03}
Typically, $\gamma_{c2}$ drops from 5-6 at low temperatures down
to $\sim$ 2 near  $T_c$.

In this paper we consider in detail the behavior of the upper
critical field for different field orientations for the case of a
\emph{dirty} two-band superconductor with weak interband
scattering. The model is based on the multiband generalization of
the quasiclassical Usadel equations.\cite{Usadel} The same model
has been used recently to describe vortex core structure in
MgB$_2$.\cite{KG} The general equations for determination of the
upper critical field within this model have been derived in recent
paper \onlinecite{Gurevich}. However, calculations in this paper
have been done only for the case of small band anisotropies. In
this paper we address the case of very high anisotropy in the
quasi-2D band, more suitable for MgB$_{2}$.

We demonstrate that the strong temperature dependence of the
$H_{c2}$-anisotropy  exists also in the dirty case and therefore
represents a general property of a two-band superconductor. The
main reason for this dependence is the strong reduction of the
in-plane upper critical field by the weak $\pi$-band in the very
narrow temperature region near $T_c$. This also leads to the
significant upward curvature of the temperature dependence of the
in-plane upper critical field near $T_c$. This behavior
illustrates breakdown of the anisotropic Ginzburg-Landau (GL)
theory for description of this superconductor. We demonstrate
that, due to the large difference between microscopic coherence
lengths in the c-direction for the two bands, the anisotropic GL
theory is applicable only within the extremely narrow temperature
range near $T_c$.

We analyze the angular dependence of the upper critical field and
show that it strongly deviates from the standard
``effective-mass'' dependence predicted by the anisotropic GL
theory. Contrary to naive expectations, these deviations are
strongest for temperatures quite close to $T_c$ (at $T \sim 0.9
T_c$) and vanish only for temperatures extremely close to $T_c$
(for $(T_c-T)/T_c \lesssim 1\%$). In the past the angular
dependence of the upper critical field have been studied in Ref.\
\onlinecite{Langmann} for a clean two-band superconductor. It was
shown that for the case of two weakly deformed spherical Fermi
surfaces with opposite anisotropies the angular dependence also
strongly deviates from the ``effective-mass'' law.

The paper is organized as follows. In section \ref{sec:model} we
present Usadel equations for a two-band superconductor and
introduce parameters relevant for MgB$_2$. In section
\ref{sec:c-axis} we derive equation for the upper critical field
in the c-direction and obtain the exact asymptotics at small and
high temperatures. In section \ref{sec:a-axis} we consider the
in-plane upper critical field. We derive general equations for
determination of this field and study solutions of these equations
in different regimes. We demonstrate that the GL result for the
in-plane $H_{c2}$ is valid only within a very narrow range of
temperatures. We also numerically calculate in-plane $H_{c2}$ and
the anisotropy parameter $\gamma_{c2}$ in the whole temperature
range. In section \ref{sec:tilted} we study the angular dependence
of the upper critical field and analyze quantitatively the
deviations from the effective-mass law.

\section{\label{sec:model}The model: Usadel equations for a two-band superconductor}

We consider a two-band superconductor with weak interband impurity
scattering and rather strong intraband scattering rates exceeding
the corresponding energy gaps (dirty limit). In this case the
quasiclassical Usadel equations \cite{Usadel} are applicable
within each band. The mixed state in this case is described by the
system of coupled Usadel equations \cite{Usadel,KG}
\begin{subequations}
\begin{align}
  \omega
F_{\alpha}\!&-\!\sum_{j}\frac{\mathcal{D}_{\alpha,j}}{2}\left[
G_{\alpha}(\nabla_{j}\!-\!\frac{2\pi
i}{\Phi_{0}}A_{j})^{2}F_{\alpha
}\!-\!F_{\alpha}\nabla_{j}^{2}G_{\alpha}\right]
\nonumber \\
&=\Delta_{\alpha}G_{\alpha},\label{UsadelFG}\\
\Delta_{\alpha}&=2\pi
T\sum_{\beta,n}\Lambda_{\alpha\beta}F_{\beta}, \label{SelfCons}
\end{align}
\end{subequations}
where $\alpha=1,2$ is the band index, $j=x,y,z$ is the coordinate
index, $\hat{\Lambda}$ is the matrix of effective coupling
constants, $\mathcal{D} _{\alpha,j}$ are diffusion constants,
which determine the coherence lengths
$\xi_{\alpha,j}=\sqrt{\mathcal{D}_{\alpha,j}/2\pi T_{c}}$,
$G_{\alpha},$ $F_{\alpha}$ and $\Delta_{\alpha}$ are normal and
anomalous Green's functions and the pair potential, respectively,
and $\omega=2\pi T(s+1/2)$ are Matsubara frequencies. Bearing in
mind the application to MgB$_{2}$, in our notations index 1
corresponds to $\sigma$-bands and index 2 to $\pi$-bands. All
bands are isotropic in the $xy$ plane, $\mathcal{D}_{\alpha
x}=\mathcal{D}_{\alpha y}$ and anisotropic in the $xz$ plane with
the anisotropy ratios $\gamma_{\alpha}=\sqrt{\mathcal{D}_{\alpha
x}/\mathcal{D}_{\alpha z}}$. The multigap Usadel equations for
general case, taking into account also interband scattering, have
been recently derived in Ref.\ \onlinecite{Gurevich}.

The selfconsistency equation can be rewritten in the form
\begin{subequations}
\label{SelfCons2}
\begin{align}
 \! W_{1}\Delta_{1}\!-W_{12}\Delta_{2}\!=&2\pi
T\!\sum_{\omega>0}\!\left(
F_{1}\!-\frac{\Delta_{1}}{\omega}\right)  \!+\Delta_{1}\ln\frac{T_{c}} {T},\\
\!-W_{21}\Delta_{1}\!+\!W_{2}\Delta_{2}\!\!=&2\pi T\!\sum_{\omega
>0}\!\left(  F_{2}\!-\frac{\Delta_{2}}{\omega}\!\right)  \!+\!\Delta_{2}\ln
\frac{T_{c}}{T},
\end{align}
with the following matrix $W_{\alpha\beta}$
\end{subequations}
\begin{align}
W_{1}  &  \!=\frac{-A+\sqrt{A^{2}+\Lambda_{12}\Lambda_{21}}}{\mathrm{Det}
},\ W_{2}\!=\frac{A+\sqrt{A^{2}+\Lambda_{12}\Lambda_{21}}}{\mathrm{Det}
},\nonumber\\
W_{12}  &  =\Lambda_{12}/\mathrm{Det},\ W_{21}=\Lambda_{21}/\mathrm{Det,}
\label{MatrixLambda}
\end{align}
$A=(\Lambda_{11}-\Lambda_{22})/2,$ $\mathrm{Det}=\Lambda_{11}\Lambda
_{22}-\Lambda_{12}\Lambda_{21}$, $W_{1}W_{2}=W_{12}W_{21}$.

The electron-phonon interaction in MgB$_{2}$ was calculated from
first principles in a number of papers.\cite{LiuPRL01,Choi,Golub}
Here we use the effective coupling constants $\Lambda_{ij}$ from
Ref.\ \onlinecite{Golub}: $\!\Lambda_{11}\!\approx0.81,\
\Lambda_{22}\!\approx0.278,\ \Lambda _{12}\!\approx0.115,\
\Lambda_{21}\!\approx0.091,$ from which we obtain values of
$W_{\alpha\beta}$ used in numerical calculations,
\begin{equation}
\!W_{1}\!\approx0.088,\ W_{2}\!\approx2.56,\ W_{12}\!\approx0.535,\ W_{21}
\!\approx0.424.
\end{equation}
The relative role of the weak band is characterized by the ratio
$S_{12}\equiv W_{1}\!/W_{2}$,\cite{Parameters} which in the case
of MgB$_{2}$ is rather small, $S_{12}\approx0.034$. This ratio
will be used below as a small parameter in our model to derive
various approximations for the upper critical field. Another
important small parameter is the ratio of diffusion coefficients
in the $\sigma$-band, $\mathcal{D}_{1z}/\mathcal{D}_{1x}$. We will
show in this paper that these two parameters, $S_{12}$ and
$\mathcal{D} _{1z} /\mathcal{D}_{1x},$ influence differently
$H_{c2}$ for parallel field at high and low temperatures thus
causing the temperature dependence of the anisotropy.

In the following we consider separately the cases when the field is parallel
and perpendicular to the ab-plane.

\section{\label{sec:c-axis}Field in the $c$-direction}

Let us first study the case when the magnetic field is oriented
along c-axis. The upper critical field is determined by the
linearized Usadel equation
\begin{equation}
\omega F_{\alpha}+\frac{\mathcal{D}_{\alpha x}}{2}\left(
-\nabla_{x} ^{2}F_{\alpha}+\left(  \frac{2\pi Hx}{\Phi_{0}}\right)
^{2}F_{\alpha}\right)
=\Delta_{\alpha} \label{Falpha_c}
\end{equation}
and selfconsistency equations (\ref{SelfCons2}). Solving these
equations, we arrive at the equation for $H_{c2}^{\perp}$ (symbol
$\perp$ denotes the field direction perpendicular to the
(\textit{ab})- plane)
\begin{equation}
\ln\frac{1}{t}-g\left(  \frac{H_{c2}^{\perp}}{tH_{1}}\right)  =-\frac
{W_{1}\left(  \ln\frac{1}{t}-g\left(  \frac{H_{c2}^{\perp}}{tH_{2}}\right)
\right)  }{W_{2}-\left(  \ln\frac{1}{t}-g\left(  \frac{H_{c2}^{\perp}}{tH_{2}
}\right)  \right)  }, \label{Eq-Hc2}
\end{equation}
where $t=T/T_{c}$,
$H_{\alpha}\equiv2T_{c}\Phi_{0}/\mathcal{D}_{\alpha x}$,
$g(x)=\psi(1/2+x)-\psi(1/2)$, and $\psi(x)$ is a digamma function.
We also obtain a relation between $\Delta_{01}$ and $\Delta_{02}$
near $H_{c2}^{\perp}$
\begin{equation}
\Delta_{02}=\frac{W_{21}\Delta_{01}}{W_{2}-\ln\frac{1}{t}+g\left(
\frac{H_{c2}^{\perp}}{tH_{2}}\right)  }.
\end{equation}
In the absence of coupling to the weak $\pi$-band ($W_{1}=0$) or
in the case of identical diffusion constants
($\mathcal{D}_{1x}=\mathcal{D}_{2x}$), the upper critical field
$H_{c2}^{s}$ is given by the standard Maki - de Gennes equation
\cite{MakiHc2,deGennesHc2}
\begin{equation}
\ln(1/t)=g\left[  H_{c2}^{s}/(tH_{1})\right]  . \label{MakiDeGennes}
\end{equation}
The well-known asymptotic solutions of this equation at low and high
temperatures are respectively
\begin{equation}
\frac{H_{c2}^{s}(t)}{H_{1}}=\left\{
\begin{array}
[c]{c}
e^{-\gamma_{E}}/4\approx0.140,\text{ }t\ll1,\\
2(1-t)/\pi^{2}\approx0.203(1-t),\text{ }1-t\ll1,
\end{array}
\right.
\end{equation}
where $\gamma_{E}\approx0.577$ is Euler constant. In the
temperature range near $T_{c}$ one can obtain from Eq.\
(\ref{Eq-Hc2}) the following simple expression for
$H_{c2}^{\perp}$ for arbitrary ratio $S_{12}\equiv W_{1}/W_{2}$:
\begin{equation}
\frac{H_{c2}^{\perp}}{H_{1}}=\frac{2\left(  1+S_{12}\right)
(1-t)}{\pi^{2}\left(  1+S_{12}\mathcal{D}_{2x}/
\mathcal{D}_{1x}\right)  }. \label{HcTc}
\end{equation}

At small temperatures, $T\ll T_{c}$, Eq.\ (\ref{Eq-Hc2}) also has
an exact solution (see also Ref.\ \onlinecite{Gurevich})
\begin{widetext}
\begin{equation}
H_{c2}(0)=H_{c2}^{s}(0)\exp\left(  -\frac{W_{1}+W_{2}-\ln\left(  r_{x}\right)
}{2}+\sqrt{\frac{\left(  W_{1} +W_{2}-\ln\left(  r_{x}\right)  \right)  ^{2}}
{4}+W_{1}\ln\left(  r_{x}\right)  }\right)  \label{Hc20Exact}
\end{equation}
\end{widetext}
with $r_{x}\equiv\mathcal{D}_{1x}/\mathcal{D}_{2x}$. For MgB$_{2}$
the parameter $W_{1}$ is small and typically the inequality
$W_{1}\ln\left( r_{x}\right)  \ll\left(  W_{2}-\ln\left(
r_{x}\right)  \right)  ^{2}/4$ is valid. In this case we can
expand Eq.\ (\ref{Hc20Exact}) with respect to $W_{1}$ and obtain a
much simpler result
\begin{equation}
H_{c2}(0)\approx H_{c2}^{s}(0)\left(  1+\frac{W_{1}\ln\left(  \mathcal{D}
_{1x}/\mathcal{D}_{2x}\right)  }{W_{2}-\ln\left(  \mathcal{D}_{1x}/\mathcal{D}
_{2x}\right)  }\right)  . \label{HcT0}
\end{equation}
The $\pi$-band strongly influences the upper critical field only if it is very
dirty, $\mathcal{D}_{2x}\ll\mathcal{D}_{1x} \exp(-W_{2})$. In this limit we
obtain \cite{GurevichError}
\[
H_{c2}(0)\approx H_{c2}^{(2)}(0)\exp\left(  -W_{2}\right)
\]
with $H_{c2}^{(2)}(0)\equiv(\exp(-\gamma_{E})/4)H_{2}$.

For the case $W_{1}\ll W_{2}$ realized in MgB$_{2}$, the upper critical field
is typically determined by the strong band (except for the limit of very small
diffusivity $\mathcal{D}_{2x}$ in the second band).
A small correction due to the weak band can be found from Eq.\
(\ref{Eq-Hc2}) using an expansion with respect to the small
parameter $S_{12}\equiv W_{1}/W_{2} $. In particular, we found
very simple expressions for the slope of $H_{c2}$ at $T_{c}$ and
$H_{c2}(0)$:
\begin{subequations}
\begin{align}
\frac{dH_{c2}}{dT}  &  \approx\frac{dH_{c2}^{s}}{dT}\left(  1+S_{12}
\frac{\mathcal{D}_{1x}-\mathcal{D}_{2x}}{D_{1x}}\right)  ,\label{Hc2cHighT}\\
H_{c2}(0)  &  \approx H_{c2}^{s}(0)\left(  1+S_{12}\ln\frac{\mathcal{D}_{1x}
}{\mathcal{D}_{2x}}\right)  . \label{Hc2cLowT}
\end{align}
The signs of the above corrections to the universal curve following from
Eq.\ (\ref{MakiDeGennes}) are positive if $D_{2x}<D_{1x}$ and negative for
$D_{2x}>D_{1x}$ .

\section{\label{sec:a-axis}Field in the $a$ -direction}

\subsection{General relations}

The upper critical field in the $a$-direction ($\equiv
y$-direction) is determined by the linear equations for the
Green's functions $F_{\alpha}$ in two bands
\end{subequations}
\begin{equation}
\omega F_{\alpha}-\frac{\mathcal{D}_{\alpha x}}{2}\nabla_{x}^{2}F_{\alpha
}+\frac{\mathcal{D}_{\alpha z}}{2}\left(  \frac{2\pi Hx}{\Phi_{0}}\right)
^{2}F_{\alpha}=\Delta_{\alpha} \label{Falpha_a}
\end{equation}
with $\omega=2\pi T(s+1/2)$ and the self-consistency conditions
(\ref{SelfCons2}). A technical difficulty of this problem is that,
due to the difference in the anisotropy factors for the two bands,
$\gamma_{\alpha}$, the harmonic oscillator operators in Eq.\
(\ref{Falpha_a}) have unmatching sets of eigenstates. We will use
an expansion with respect to the eigenfunctions (Landau levels) of
the strong [1st] band, $\Psi_{n}(x)$, which are defined as
solutions of the oscillator equation
\begin{equation}
\frac{\mathcal{D}_{1z}}{2}\left(  \frac{2\pi H}{\Phi_{0}}\right)
^{2}
x^{2}\Psi_{n}-\frac{\mathcal{D}_{1x}}{2}\nabla_{x}^{2}\Psi_{n}=\varepsilon
_{n}\Psi_{n}.
\end{equation}
In particular, the eigenvalues $\varepsilon_{n}$ and ground state
eigenfunction are given by
\begin{align}
\varepsilon_{n}  &  =\sqrt{\mathcal{D}_{1z}\mathcal{D}_{1x}}\frac{2\pi H}
{\Phi_{0}}\left(  n+1/2\right)  ,\ \\
\Psi_{0}(x)  &  =\left(  \frac{2H}{\gamma_1\Phi_{0}}\right)
^{1/4} \exp\left(  -\frac{\pi Hx^{2}}{\gamma_1\Phi_{0}}\right)  ,
\label{eigenstates}
\end{align}
where $\gamma_{\alpha}=\sqrt{\mathcal{D}_{\alpha x}/\mathcal{D}_{\alpha z}}$
are the band anisotropies. In the case of MgB$_{2}$ the first band is
quasi-two-dimensional, i.e.,$\ \gamma_{1}\gg1,\gamma_{2}$. Substituting
expansions
\[
\Delta_{\alpha}(x)=\sum_{n}\Delta_{\alpha,n}\Psi_{n}(x);\ F_{\alpha}=\sum
_{n}F_{\alpha,n}\Psi_{n}(x)
\]
into Eq.\ (\ref{Falpha_a}), we obtain
\begin{subequations}
\begin{align}
F_{1,n}  &  =\frac{\Delta_{1,n}}{\omega+\varepsilon_{n}},\\
\omega F_{2,n}+\sum_{m=0}^{\infty}\epsilon_{nm}F_{2,m}  &  =\Delta_{2,n}
\end{align}
with $\epsilon_{nm}=\left\langle \frac{\mathcal{D}_{2z}}{2}\left(
\frac{2\pi H}{\Phi_{0}}\right)
^{2}x^{2}-\frac{\mathcal{D}_{2x}}{2}\nabla_{x} ^{2}\right\rangle
_{nm}$. The only nonzero matrix elements $\epsilon_{nm}$ are at
$m=n$ and $m=n\pm2$:
\end{subequations}
\begin{subequations}
\begin{align}
\epsilon_{nn}  &  =\frac{\pi H}{\Phi_{0}}\sqrt{\mathcal{D}_{2x}\mathcal{D}
_{2z}}\left(  n+\frac{1}{2}\right)  \left(  \frac{\gamma_{2}}{\gamma_{1}
}+\frac{\gamma_{1}}{\gamma_{2}}\right)  ,\\
\epsilon_{n-2,n}  &  =\epsilon_{n,n-2}=\frac{\pi
H\sqrt{\mathcal{D}_{2x}\mathcal{D}_{2z}}}{\Phi_{0}}\frac
{\sqrt{n(n-1)}}{2}\left(  \frac
{\gamma_{2}}{\gamma_{1}}-\frac{\gamma_{1}}{\gamma_{2}}\right)  .
\end{align}
Neglecting the small ratio $\gamma_{1}/\gamma_{2}$ in comparison
with $\gamma _{2}/\gamma_{1}$ we obtain
\end{subequations}
\begin{subequations}
\label{MatrixElemApprox}
\begin{align}
\epsilon_{nn}  &  \approx\left(  n+\frac{1}{2}\right)  w_{2},\\
\epsilon_{n-2,n}  &  =\epsilon_{n,n-2}\approx-\frac{\sqrt{n(n-1)}}{2}w_{2}
\end{align}
with $w_{2}\equiv(\pi H/\Phi_{0})\mathcal{D}_{2z}\gamma_{1}$. This
approximation for the matrix elements is equivalent to the local
approximation for the $F$-function in the $\pi$-band described in
Appendix \ref{App:Local}. Therefore, we can rewrite the equation
for $F_{2n}$ as
\end{subequations}
\begin{equation}
\omega F_{2n}+\epsilon_{n,n-2}F_{2,n-2}+\epsilon_{n,n}F_{2,n}+\epsilon
_{n,n+2}F_{2,n}=\Delta_{2,n}.
\end{equation}
At $n=0$ the term $\epsilon_{n,n-2}F_{2,n-2}$ has to be skipped. This means
that even Landau levels, $n=2i$, do not mix with the odd Landau level,
$n=2i+1$. For calculation of the upper critical field it is sufficient to
consider only even Landau levels. The self-consistency equations in terms of
the expansion coefficients are given by \begin{widetext}
\begin{subequations}
\begin{align}
W_{1}\Delta_{1,n}-W_{12}\Delta_{2,n} &  =2\pi T\sum_{\omega>0}\left(  \frac
{1}{\omega+\varepsilon_{n}}-\frac{1}{\omega}\right)  \Delta_{1,n}+\Delta
_{1,n}\ln\frac{1}{t},\\
-W_{21}\Delta_{1,n}+W_{2}\Delta_{2,n} &  =2\pi
T\sum_{\omega>0}\left( F_{2,n}-\frac{\Delta_{2,n}}{\omega}\right)
+\Delta_{2,n}\ln\frac{1}{t} .
\end{align}
\end{subequations}
To simplify further analysis we introduce the reduced variables
\[
z=\omega/w_{2}=t_{2}(s+1/2),\ \tilde{F}_{2,i}=w_{2}F_{2,2i}
\]
with $t_{2}\equiv2\pi
T/w_{2}\equiv2\Phi_{0}T/(H\mathcal{D}_{2z}\gamma_{1})$ and $s$ is
the Matsubara index. Then equations for $\tilde{F}_{2,i}(z)$ are
given by
\begin{subequations}
\label{F2i}
\begin{align}
\left(  z+\frac{1}{2}\right)  \tilde{F}_{2,0}-(1/\sqrt{2})\tilde{F}_{2,1} &
=\Delta_{2,0},\\
-\sqrt{i(i-1/2)}\tilde{F}_{2,i-1}+\left(  z+2i+1/2\right)
\tilde{F} _{2,i}-\sqrt{(i+1)(i+1/2)}\tilde{F}_{2,i+1} &
=\Delta_{2,2i}.
\end{align}
\end{subequations}
The formal solution of Eq.\ (\ref{F2i}) is given by
\[
\tilde{F}_{2,i}(z)=\sum_{j=0}^{\infty}A_{i,j}(z)\Delta_{2,2j},
\]
where the matrix $A_{i,j}(z)$ is defined as solution of equations
\begin{subequations}
\label{A0j}
\begin{align}
\left(  z+\frac{1}{2}\right)  A_{0,j}-\sqrt{1/2}A_{1,j} &  =\delta_{0,j},\\
-\sqrt{i(i-1/2)}A_{i-1,j}+\left(  z+2i+1/2\right)  A_{i,j}-\sqrt
{(i+1)(i+1/2)}A_{i+1,j} &  =\delta_{i,j}.
\end{align}
\end{subequations}
\end{widetext}Using this solution we represent the self-consistency conditions
for even Landau levels in the form
\begin{subequations}
\begin{align}
W_{1}\Delta_{1,2i}\!-W_{12}\Delta_{2,2i}  &
\!=\Delta_{1,2i}\left( \ln\frac {1}{t}-g\left(
\frac{H(4i+1)}{tH_{1}^{\parallel}}\right) \right)
,\label{SelfConsReducedA}\\
-W_{21}\Delta_{1,2i}\!+W_{2}\Delta_{2,2i}  &  \!=\sum_{j=0}^{\infty}U_{i,j}
(t_{2})\Delta_{2,2j}+\Delta_{2,2i}\ln\frac{1}{t} \label{SelfConsReducedB}
\end{align}
with
\end{subequations}
\begin{equation}
H_{1}^{\parallel}\equiv\frac{2T_{c}\Phi_{0}}{\sqrt{\mathcal{D}_{1z}
\mathcal{D}_{1x}}}, \label{H1eff}
\end{equation}
where symbol $\parallel$ denotes the field direction parallel to the
(\textit{ab})- plane, and
\begin{equation}
U_{i,j}(t_{2})\!=\!t_{2}\!\sum_{s=0}^{\infty}\left(
A_{i,j}\left[t_{2}(s+1/2)\right]\!-\!\frac
{\delta_{i,j}}{t_{2}(s+1/2)}\right)  . \label{Uij}
\end{equation}
We again used notations $t=T/T_{c}$ and $g(x)\equiv\psi\left(
1/2+x\right) -\psi(1/2)$. We show in Appendix \ref{App:Local} that
$U_{i,j}(t_{2})$ can also be related with the oscillator matrix
element of the function $g\left[ x^{2}/t_{2}\right]  $
\begin{align*}
U_{i,j}(t_{2})  &  =\int_{-\infty}^{\infty}dx\psi_{2i}(x)\psi_{2j}(x)g\left[
x^{2}/t_{2}\right]  ,\\
\psi_{n}(x)  &  =\frac{\exp\left(  -x^{2}/2\right)
H_{n}(x)}{\pi^{1/4} \sqrt{2^{n}n!}},
\end{align*}
where $H_{n}(x)$ are Hermite polynomials.

\subsection{Temperatures not close to $T_c$. High-field approximation in the $\pi
$-band}

The overall behavior is determined by the value of dimensionless parameter
$t_{2}$, which depends on field and temperature. To evaluate this parameter we
represent it in the form
\begin{equation}
t_{2}=\frac{\mathcal{D}_{1z}}{\mathcal{D}_{2z}}\frac{tH_{1}^{\parallel}}{H}.
\label{t2}
\end{equation}
Because $\mathcal{D}_{1z}\ll\mathcal{D}_{2z}$ and at low
temperatures $H\lesssim H_{1}^{\parallel}$, the parameter $t_{2}$
is much smaller than unity almost in the whole temperature range
except a very narrow region near $T_{c}$. The parameter $t_{2}$
becomes of the order of one only at
$(T_{c}-T)/T_{c}\sim\mathcal{D}_{1z}/\mathcal{D}_{2z}\ll1$.
Outside this region one can replace summation with respect to the
Matsubara index $s$ in Eq.\ (\ref{Uij}) by integration, which
allows us to reduce it to the following form
\[
U_{i,j}(t_{2})\approx f_{i,j}+\left(  \ln t_{2}-\gamma_{E}-2\ln2\right)
\delta_{i,j},
\]
where
\begin{align}
f_{i,j}  &  =\int_{0}^{\infty}dz\left(  A_{i,j}(z)-\frac{\delta_{i,j}}
{z+1}\right) \label{fij}\\
&  =-4\int_{0}^{\infty}dx\psi_{2i}(x)\psi_{2j}(x)\ln(x)
\end{align}
is the universal matrix of constants (in particular,
$f_{0,0}=\gamma_{E} +2\ln2\approx1.96$). Using this
representation, we transform Eq.\ (\ref{SelfConsReducedB}) to the
form
\begin{widetext}
\begin{equation}
-W_{21}\Delta_{1,2i}+W_{2}\Delta_{2,2i}=\sum_{j=0}^{\infty}f_{i,j}
\Delta_{2,2j}-\left(  \ln\left(  \frac{H}{H_{1}^{\parallel}}\frac
{\mathcal{D}_{2z}}{\mathcal{D}_{1z}}\right)
+\gamma_{E}+2\ln2\right) \Delta_{2,2i}. \label{SelfConsFinal}
\end{equation}
We refer to this approximation as the high-field regime in the
$\pi$-band. The last equation in combination with Eq.\
(\ref{SelfConsReducedA} ) determines the upper critical field
along $a$-direction within the "high-field in the $\pi$-band"
regime, at $(T_{c}-T)/T_{c}\gg\mathcal{D} _{1z}/\mathcal{D}_{2z}$.
Note that in this approximation the temperature dependence exists
only in Eq.\ (\ref{SelfConsReducedA}). Therefore, once computed,
matrix $f_{i,j}$ allows us to calculate the temperature dependence
of $H_{c2}^{\parallel}$ in a wide temperature range.

Excluding $\Delta_{1,2i}$
\begin{equation}
\Delta_{1,2i}=\frac{W_{12}}{W_{1}-\left[  \ln\frac{1}{t}-g\left(
\frac{H(4i+1)}{tH_{1}^{\parallel}}\right)  \right] }\Delta_{2,2i},
\end{equation}
we also derive equations containing only $\Delta_{2,2i}$
\begin{equation}
\left(  \ln\left(
\frac{H}{H_{1}^{\parallel}}\frac{\mathcal{D}_{2z}
}{\mathcal{D}_{1z}}\right)  +\gamma_{E}+2\ln2+\frac{-W_{2}\left(
\ln\frac {1}{t}-g\left(  \frac{H(4i+1)}{tH_{1}^{\parallel}}\right)
\right)  } {W_{1}-\left(  \ln\frac{1}{t}-g\left(
\frac{H(4i+1)}{tH_{1}^{\parallel} }\right)  \right)  }\right)
\Delta_{2,2i}-\sum_{j=0}^{\infty}f_{i,j} \Delta_{2,2j}=0.
\end{equation}
\end{widetext}The upper critical field $H=H_{c2}^{\parallel}$ is given by the
maximum root of the determinant of this linear system. An approximate solution
can be obtained neglecting coupling to the higher Landau levels in the
self-consistency equations leading to the following equation for
$H_{c2}^{\parallel}$
\begin{equation}
\ln\frac{1}{t}-g\left(
\frac{H_{c2}^{\parallel}}{tH_{1}^{\parallel}}\right)
=\frac{W_{1}\ln\left(
\frac{H_{c2}^{\parallel}}{H_{1}^{\parallel}}
\frac{\mathcal{D}_{2z}}{\mathcal{D}_{1z}}\right)
}{W_{2}+\ln\left(
\frac{H_{c2}^{\parallel}}{H_{1}^{\parallel}}\frac{\mathcal{D}_{2z}
}{\mathcal{D}_{1z}}\right)  }. \label{HaT0}
\end{equation}
Since $W_{1}\ll1$, the right hand side of Eq.(\ref{HaT0}) is
small. As a result, in the limit of small $t_{2}$ the parallel
critical field is close to the solution of the Maki - de Gennes
equation (\ref{MakiDeGennes}) with the effective parameter $H_{1}$
replaced by $H_{1}^{\parallel}$ from Eq.\ (\ref{H1eff}). A small
correction from the weak band can be estimated at low temperatures
\begin{equation}
H_{c2}^{\parallel}(0)\approx H_{c2}^{s\parallel}(0)\left(  1-\frac
{W_{1}\left(  \ln\left(  \mathcal{D}_{2z}/\mathcal{D}_{1z}\right)
-1.96\right)  }{W_{2}+\ln\left(
\mathcal{D}_{2z}/\mathcal{D}_{1z}\right) -1.96}\right).
\label{Hc2aT0}
\end{equation}
with $H_{c2}^{s\parallel}(0)=(\exp(-\gamma_{E})/4)H_{1}^{\parallel}$.

Combining Eqs.\ (\ref{HcT0}) and (\ref{Hc2aT0}) we obtain an
estimate for the anisotropy factor
$\gamma_{c2}(T)=H_{c2}^{\parallel}(T)/H_{c2}^{\perp}(T)$ at low
temperatures \begin{widetext}
\begin{equation}
\gamma_{c2}(0)\approx\gamma_1\left(  1+\frac{W_{1}\ln\left(
\mathcal{D}_{2x}/\mathcal{D}_{1x}\right)  }{W_{2}+\ln\left(
\mathcal{D} _{2x}/\mathcal{D}_{1x}\right)  }-\frac{W_{1}\left(
\ln\left(  \mathcal{D} _{2z}/\mathcal{D}_{1z}\right)  -1.96\right)
}{W_{2}+\ln\left(  \mathcal{D} _{2z}/\mathcal{D}_{1z}\right)
-1.96}\right) . \label{gamma_corr}
\end{equation}
As follows from this equation,
the anisotropy of $H_{c2}$ at $T=0$ is very close to the anisotropy of the
first band
\begin{equation}
\gamma_{c2}(0)\approx\gamma_{1}\equiv\sqrt{\mathcal{D}_{1x}/\mathcal{D}_{1z}}.
\label{gamma0}
\end{equation}
To estimate the ratio
$\mathcal{D}_{1x}/\mathcal{D}_{1z}=v_{F1x}^{2}\tau
_{1z}/v_{F1z}^{2}\tau_{1x}$ for MgB$_{2}$, we take
$v_{F1x}^{2}/v_{F1z} ^{2}\approx40$ provided in Ref.\
\onlinecite{Brink} and assume isotropic scattering
$\tau_{1x}\approx\tau_{1z}$. This gives $\sqrt{\mathcal{D}
_{1x}/\mathcal{D}_{1z}}\approx\sqrt{40}\approx6.3$, which is
consistent with the experimental data on the $H_{c2}$ anisotropy
in MgB$_{2}$ single
crystals.\cite{Sologub,Budko,Angst,Eltsev,Lyard}

\subsection{Ginzburg-Landau region}

In the close vicinity of $T_{c}$ (exact criterion will be
established below) one can solve Eq. (\ref{Falpha_a}) using the
gradient expansion
\[
F_{\alpha}\approx\frac{\Delta_{\alpha}}{\omega}-\frac{1}{\omega^{2}}\left(
-\frac{\mathcal{D}_{\alpha
x}}{2}\nabla_{x}^{2}\Delta_{\alpha}+\frac {\mathcal{D}_{\alpha
z}}{2}\left(  \frac{2\pi Hx}{\Phi_{0}}\right)  ^{2}
\Delta_{\alpha}\right).
\]
Substituting this expansion into the self-consistency conditions
and using relation $2\pi T\sum_{\omega>0}(1/\omega^{2})=\pi/4T$,
we obtain
\begin{subequations}
\begin{align}
W_{1}\Delta_{1}-W_{12}\Delta_{2}  &  =-\left(  -\mathcal{\xi}_{1x}^{2}
\nabla_{x}^{2}\Delta_{1}+\mathcal{\xi}_{1z}^{2}\left(  \frac{2\pi Hx}{\Phi
_{0}}\right)  ^{2}\Delta_{1}\right)  +\Delta_{1}\ln\frac{1}{t}\label{GL1}\\
-W_{21}\Delta_{1}+W_{2}\Delta_{2}  &  =-\left(  -\mathcal{\xi}_{2x}^{2}
\nabla_{x}^{2}\Delta_{2}+\mathcal{\xi}_{2z}^{2}\left(  \frac{2\pi Hx}{\Phi
_{0}}\right)  ^{2}\Delta_{2}\right)  +\Delta_{2}\ln\frac{1}{t} \label{GL2}
\end{align}
\end{subequations}
\end{widetext}
with $\mathcal{\xi}_{\alpha i}^{2}\equiv\pi\mathcal{D}_{\alpha
i}/(8T)$. Near $T_{c}$ we can look for solution for $\Delta_{2}$
in the form
\[
\Delta_{2}\approx\frac{W_{21}}{W_{2}}\Delta_{1}+\delta_{2},
\]
where $\delta_{2}$ is a small correction, for which we obtain from
Eq.\ (\ref{GL2})
\[
-W_{12}\delta_{2}\approx-\left(
-\mathcal{\xi}_{1x}^{2}\nabla_{x}^{2}
\Delta_{1}+\mathcal{\xi}_{1z}^{2}\left(  \frac{2\pi
Hx}{\Phi_{0}}\right) ^{2}\Delta_{1}\right)
+\Delta_{1}\ln\frac{1}{t}.
\]
Substituting this result into Eq.\ (\ref{GL1}) we obtain the
linear Ginzburg-Landau (GL) equation for $\Delta_{1}$
\begin{equation}
-\xi_x^2\nabla_{x}^{2}\Delta_{1}+\xi_z^2\left( \frac{2\pi
Hx}{\Phi_{0}}\right) ^{2}\Delta_{1}-\Delta_{1}\ln\frac{1}{t}=0,
\end{equation}
in which the averaged coherence lengths, $\xi_i$ with $i=x,z$, are
defined as
\begin{equation}
\xi_{i}=\sqrt{\frac{\xi_{1i}^{2}+S_{12}\xi_{2i}^{2}}{1+S_{12}}}.
\label{xi_i}
\end{equation}
From this equation we immediately obtain the usual GL result for
the upper critical field at $T\rightarrow T_{c}$
\begin{equation}
H_{c2}=\frac{\Phi_{0}(1-t)}{2\pi\xi_{x}\xi_{z}},
\end{equation}
For comparison with numerical results at lower temperature we also
provide $H_{c2}^{\parallel}$ in units of $H_{1}^{\parallel}$
\begin{align}
\frac{H_{c2}^{\parallel}}{H_{1}^{\parallel}}  &
=\frac{2\sqrt{\mathcal{D} _{1x}\mathcal{D}_{1z}}\left(
1+S_{12}\right)  (1-t)}{\pi ^{2}\sqrt{\left(
\mathcal{D}_{1x}+S_{12}\mathcal{D}_{2x}\right)
\left(  \mathcal{D}_{1z}+S_{12}\mathcal{D}_{2z}\right)  }
}\label{HparTc}\\
&  \approx\frac{2}{\pi^{2}\sqrt{1+S_{12}\mathcal{D}_{2z}
/\mathcal{D}_{1z}}}(1-t)\nonumber
\end{align}
for $W_{1}\ll W_{2}$ and $\mathcal{D}_{1x}\sim\mathcal{D}_{2x}$.

Due to the strong inequality
$\mathcal{D}_{2z}\gg\mathcal{D}_{1z}$, in the vicinity of $T_{c}$
the three-dimensional band strongly reduces the upper critical
field. This reduction leads to a strong temperature dependence of
the $H_{c2}$ anisotropy, $\gamma_{c2}$.

Let us compare anisotropy parameters at low $T$ and near $T_{c}$.
According to Eq.\ (\ref{gamma0}), the anisotropy of $H_{c2}$ at
low temperatures is close to the anisotropy of the $\sigma$-band,
$\gamma_{c2}(0)\approx\sqrt{\mathcal{D} _{1x}/\mathcal{D}_{1z}}$,
while the anisotropy ratio near $T_{c}$ follows from Eqs.\
(\ref{HcTc}) and (\ref{HparTc})
\begin{align}
\gamma_{c2}(T_{c})\equiv \gamma_{GL}&
=\gamma_{1}\sqrt{\frac{1+S_{12}\mathcal{D}_{2x}/\mathcal{D}_{1x}}{1+S_{12}
\mathcal{D}_{2z}/\mathcal{D}_{1z}}}\\
&
\approx\frac{\gamma_{1}}{\sqrt{1+S_{12}\mathcal{D}_{2z}/\mathcal{D}_{1z}}}.
\nonumber
\end{align}
Thus the ratio $\gamma_{c2}(0)/\gamma_{c2}(T_{c})$ is roughly given by
\begin{equation}
\frac{\gamma_{c2}(0)}{\gamma_{c2}(T_{c})}\approx\sqrt{1+S_{12}\mathcal{D}_{2z}/\mathcal{D}_{1z}}. \label{gamma_ratio}
\end{equation}
The larger is the ratio of transport constants, $\mathcal{D}_{1z}
/\mathcal{D}_{2z},$ the stronger is the suppression of
$\gamma_{c2}(T)$ with increasing temperature.

We obtain now the applicability criterion for the GL expansion. Typical scales
of the order parameter variation near $T_{c}$ are given by the GL coherence
lengths $\xi_{i}^{GL}(T)=\xi_{i}/\sqrt{1-t}$, with $i=x,y$ and $\xi_{i}$ given
by Eq.\ (\ref{xi_i}). The GL expansion is valid until the GL coherence lengths
are larger than the corresponding microscopic coherence lengths in both bands,
$\xi_{i}^{GL}(T)>\xi_{\alpha,i}$. Because of the strong inequality $\xi
_{1,z}\ll\xi_{2,z}$, the most sensitive condition is
\begin{equation}
\xi_{z}^{GL}(T)>\xi_{2,z},
\end{equation}
leading to the following condition for the GL temperature range
\begin{equation}
\frac{T_{c}-T}{T_{c}}<\max\left(
\frac{\xi_{1z}^{2}}{\xi_{2z}^{2}},S_{12}\right)  .
\end{equation}
Because $\xi_{1z}\ll\xi_{2z}$ and $S_{12}\ll 1$, the applicability
of the GL approach is limited to an extremely narrow temperature
range near $T_{c}$, i.e., the situation is very different from
usual single-band superconductors. The comparison of the GL
asymptotic and with the exact solution is shown in Fig.\
\ref{Fig-Hc2a}, where the narrowness of the GL region is
demonstrated in the inset.

\subsection{\label{Sec:num} Numerical solution in the whole temperature
range.}

\begin{figure}[ptb]
\begin{center}
\includegraphics[clip, width=3.4in ]{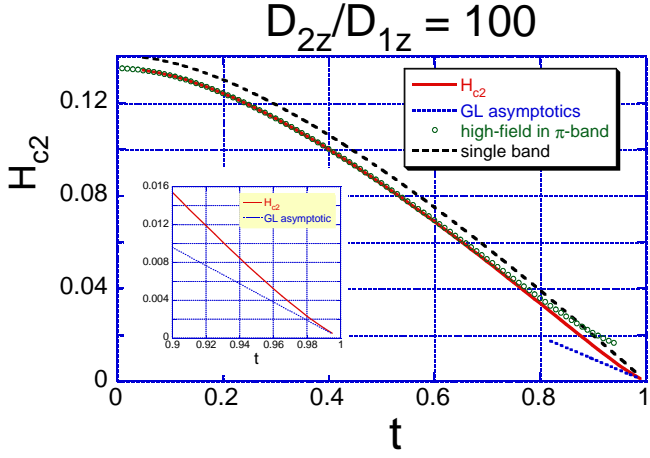}
\end{center}
\par
\caption{Temperature dependence of the upper critical field in the
a-direction normalized to $H_{1}^{\parallel}$ defined in Eq.\
(\ref{H1eff}). Single-band curve is normalized to the
corresponding scale $H_{1}$. Solid circles show the dependence
obtained within the "high-field in the $\pi$-band" approximation
(Eqs.\ (\ref{fij}) and (\ref{SelfConsFinal})). Inset: comparison
between the exact solution and the
GL asymptotics Eq.\ (\ref{HparTc}). }
\label{Fig-Hc2a}
\end{figure}

In the whole temperature range, for an arbitrary value of the
parameter $t_{2}$, the problem can be solved numerically. The
solution consists of three steps: (i) the matrix $A_{i,j}(z)$ has
to be found from Eqs.\ (\ref{A0j}) for the series of reduced
Matsubara frequencies $z=t_{2}(s+1/2)$, (ii) the matrix $U_{i,j}$
has to be computed by summation over Matsubara indices $s$
(\ref{Uij}) and (iii)the upper critical field has to be found as
the maximum root of the determinant of the linear system
represented by Eqs.\ (\ref{SelfConsReducedA}) and
(\ref{SelfConsReducedB}).
Due to fast decrease of the nondiagonal matrix elements $U_{i,j}$
for $\left\vert i-j\right\vert \gg1$ , sufficient accuracy is
achieved for dimension of the matrix less than $30$. The result of
calculation of the parallel upper critical field is shown in Fig.\
\ref{Fig-Hc2a} where the ratio
$\mathcal{D}_{2z}/\mathcal{D}_{1z}=100$ relevant to MgB$_{2}$ was
used. Note that when plotted in reduced units, the deviations of
both ratios $H_{c2}^{\parallel}/H_{1}^{\parallel}$ and
$H_{c2}^{\perp}/H_{1}^{\perp}$ from the universal single band
curve are small (except from the region near $T_{c} $, in the GL
region), in accordance with the above discussion. However, one
should keep in mind the large difference in magnitudes of the
characteristic scales $H_{1}^{\parallel}$ and $H_{1}^{\perp}$.

\begin{figure}[ptb]
\begin{center}
\includegraphics[clip, width=3.4in ]{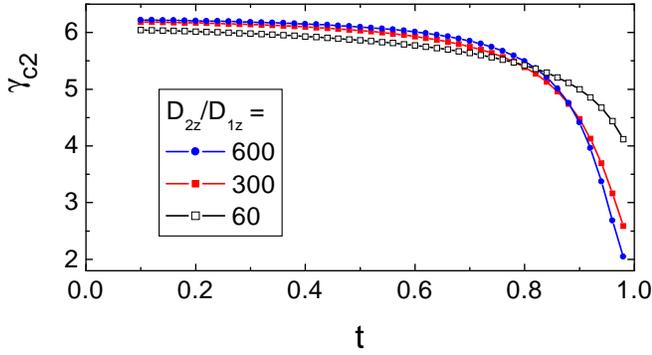}
\end{center}
\par
\caption{Temperature dependence of the anisotropy of the upper critical field
for several ratios $\mathcal{D}_{2z}/\mathcal{D}_{1z}$.}
\label{Fig-GammaT}
\end{figure}

Numerically calculated temperature dependence of the anisotropy
factor for several ratios $\mathcal{D}_{2z}/\mathcal{D}_{1z}$ is
shown in Fig.\ \ref{Fig-GammaT}. The anisotropy ratio drops with
the increase of temperature, in accordance with the estimate
(\ref{gamma_ratio}). This result agrees qualitatively with recent
measurements of temperature-dependent anisotropy in
MgB$_{2}$.\cite{Sologub,Budko,Angst,Eltsev,Lyard} In experiment
the change in anisotropy typically is distributed over wider
temperature range than it is suggested by the theory.

\section{\label{sec:tilted}Tilted fields}

The upper critical field for magnetic field tilted at angle $\theta$ with
respect to $z$ axis in $(zy)$ plane is determined by the coupled linear
equations for the Green's functions $F_{\alpha}$ in two bands
\begin{equation}
\omega F_{\alpha}-\frac{\mathcal{D}_{\alpha x}}{2}\nabla_{x}^{2}F_{\alpha
}+\frac{\mathcal{D}_{\alpha}(\theta)}{2}\left(  \frac{2\pi Hx}{\Phi_{0}}
\right)  ^{2}F_{\alpha}=\Delta_{\alpha} \label{Falpha_tilt}
\end{equation}
with
\begin{equation}
\mathcal{D}_{\alpha}(\theta)=\mathcal{D}_{\alpha x}\cos^{2}\theta+
\mathcal{D}_{\alpha z}\sin^{2}\theta\label{DiffAngle}
\end{equation}
and the self-consistency conditions (\ref{SelfCons2}).

Therefore the $H_{c2}$-problem of the upper critical field in
tilted field reduces to the in-plane $H_{c2}$-problem by
substitution $\mathcal{D}_{\alpha
z}\rightarrow\mathcal{D}_{\alpha}(\theta)$. It is convenient to
introduce the angular-dependent anisotropy parameters
\begin{equation}
\gamma_{\alpha}(\theta)\equiv\sqrt{\frac{\mathcal{D}_{\alpha x}}
{\mathcal{D}_{\alpha}(\theta)}}=\frac{\gamma_{\alpha}}{\sqrt{\gamma_{\alpha
}^{2}\cos^{2}\theta+\sin^{2}\theta}}. \label{AnisAngle}
\end{equation}
Such defined anisotropy parameters vary from $1$ to $\gamma_{\alpha}$ when
angle varies from $0$ to $\pi/2$.

Following the route of the previous Section, we again use
expansion with respect to the Landau levels of the strong band,
defined by Eq. (\ref{eigenstates}) with
$\mathcal{D}_{1z}\rightarrow\mathcal{D}_{1}(\theta)$. The
$F$-function of the strong band is given by
\[
F_{1,n}=\frac{\Delta_{1,n}}{\omega+\varepsilon_{n}(\theta)}
\]
with the eigenvalue
\begin{align*}
\varepsilon_{n}(\theta)  &  =2\pi T\frac{H}{tH_{1}(\theta)}(2n+1),\ \\
H_{1}(\theta)  &  =\frac{2T_{c}\Phi_{0}}{\sqrt{\mathcal{D}_{1}(\theta
)\mathcal{D}_{1x}}}.
\end{align*}
The matrix elements for the harmonic oscillator operator of the weak band are
given by \begin{widetext}
\begin{align*}
\epsilon_{nn}  &  =\frac{\pi H}{\Phi_{0}}\mathcal{D}_{2}(\theta)\gamma
_{1}(\theta)\left(  n+\frac{1}{2}\right)  \left(  1+\left(  \frac{\gamma
_{2}(\theta)}{\gamma_{1}(\theta)}\right)  ^{2}\right)  =\frac{2\pi T}{
t_{2}(\theta)}(1+\alpha_{\gamma}(\theta))\left(  n+\frac{1}{2}\right) \\
\epsilon_{n,n-2}  &  =-\frac{\pi
H}{\Phi_{0}}\frac{\sqrt{n(n-1)}}{2}\mathcal{D
}_{2}(\theta)\gamma_{1}(\theta)\left(  1-\left(
\frac{\gamma_{2}(\theta) }{\gamma_{1}(\theta)}\right)  ^{2}\right)
=-\frac{2\pi T}{t_{2}(\theta)}
(1-\alpha_{\gamma}(\theta))\frac{\sqrt{n(n-1)}}{2}
\end{align*}
with
\begin{align*}
t_{2}(\theta)  &  =\frac{2T\Phi_{0}}{H\mathcal{D}_{2}(\theta)\gamma_{1}
(\theta)}=\frac{2T\Phi_{0}\sqrt{\cos^{2}\theta+\gamma_{1}^{-2}\sin^{2}\theta}
}{H\mathcal{D}_{2x}\left(  \cos^{2}\theta+\gamma_{2}^{-2}\sin^{2}
\theta\right)  },\\
\alpha_{\gamma}(\theta)  &  =\left(  \frac{\gamma_{2}(\theta)}{\gamma
_{1}(\theta)}\right)  ^{2}=\frac{1+\gamma_{1}^{-2}\tan^{2}\theta}{1+\gamma
_{2}^{-2}\tan^{2}\theta}.
\end{align*}
Note that at arbitrary angle we can not use inequality $\gamma_{2}
(\theta)/\gamma_{1}(\theta)\ll1$ any more. The system of equations
for the reduced $F$-function at even Landau levels,
$\tilde{F}_{2,n}=(2\pi T/t_{2}(\theta))F_{2,2i}$, at arbitrary
tilt angle is given by
\begin{equation}
-(1-\alpha_{\gamma})\sqrt{i(i-1/2)}\tilde{F}_{2,i-1}+\left(
z+(1+\alpha _{\gamma})\left(  2i+\frac{1}{2}\right)  \right)
\tilde{F}_{2,i}
-(1-\alpha_{\gamma})\sqrt{(i+1/2)(i+1)}\tilde{F}_{2,i+1}=\Delta_{2,2i}
\label{F-eq-tilted}
\end{equation}
\end{widetext}with $z=t_{2}(\theta)(s+1/2)$.

At small tilt angles, $\theta\ll1$, one can solve Eq.\ (\ref{F-eq-tilted})
using perturbation theory with respect to $\theta^{2}$. The quadratic angular
correction can be obtained neglecting coupling to the higher Landau level.
This leads to equation similar to Eq.\ (\ref{Eq-Hc2}) with replacements
\begin{align*}
H_{1}  &  \rightarrow H_{1}(\theta)=\frac{2T_{c}\Phi_{0}}{\sqrt{\mathcal{D}
_{1x}\mathcal{D}_{1}(\theta)}},\\
H_{2}  &  \rightarrow
H_{2}(\theta)=\frac{4T_{c}\Phi_{0}}{\mathcal{D}
_{2}(\theta)\gamma_{1}(\theta)(1+\alpha_{\gamma})}.
\end{align*}
At small angles we obtain quadratic in $\theta$ corrections to typical fields
\begin{align*}
H_{1}(\theta)  &  \approx H_{1}\left(  1+(1-\gamma_{1}^{-2})\frac{\theta^{2}
}{2}\right)  ,\\
H_{2}(\theta)  &  \approx H_{2}\left(
1+(1-\gamma_{2}^{-2})\frac{\theta^{2} }{2}\right).
\end{align*}
At low temperature one can derive an exact formula for small-angle
correction
\begin{widetext}
\begin{equation}
\frac{H_{c2}(\theta)-H_{c2}(0)}{H_{c2}(0)}\approx\frac{\theta^{2}}{2}\left(
1-\frac{1}{2}\left(  \gamma_{1}^{-2}+\gamma_{2}^{-2}-\frac{\left(
\gamma _{2}^{-2}-\gamma_{1}^{-2}\right)  (W_{2}-W_{1}-\ln
r_{x})}{\sqrt{\left( W_{2}+W_{1}-\ln r_{x}\right)  ^{2}+4W_{1}\ln
r_{x}}}\right)  \right) \label{Small-Tilt-Exact}
\end{equation}
with $r_{x}\equiv\mathcal{D}_{1x}/\mathcal{D}_{2x}$. In the case
of small correction from the weak band, $4W_{1}\ln r_{x}\ll\left(
W_{2}-\ln r_{x}\right)  ^{2}$, we obtain a simpler formula for
$\theta\ll1$
\begin{equation}
\frac{H_{c2}(\theta)-H_{c2}(0)}{H_{c2}(0)}\approx\frac{\theta^{2}}{2}\left(
1-\gamma_{1}^{-2}+\frac{W_{2}W_{1}\left(
\gamma_{1}^{-2}-\gamma_{2} ^{-2}\right)  }{\left(  W_{2}-\ln
r_{x}\right)  ^{2}}\right)  .
\end{equation}
For parameters of MgB$_{2}$ this formula gives an estimate almost identical to
the exact result.

At large tilt angles, $\cos\theta\ll1$, inequality
$\gamma_{2}(\theta )\ll\gamma_{1}(\theta)$ is restored and we can
utilize the approximations used for the case of in-plane field. In
particular, at low temperatures the approximate angular dependence
is given by a formula similar to Eq.\ (\ref{Hc2aT0}),
\begin{equation}
H_{c2}(\theta)\approx\frac{H_{c2}^{(1)}(0)}{\sqrt{\cos^{2}\theta+\gamma
_{1}^{-2}\sin^{2}\theta}}\left(  1-\frac{W_{1}\left(
\ln\frac{\mathcal{D}
_{2z}}{\mathcal{D}_{1x}\cot^{2}\theta+\mathcal{D}_{1z}}-1.96\right)
} {W_{2}+\ln\frac{\mathcal{D}_{2z}}{\mathcal{D}_{1x}\cot^{2}\theta
+\mathcal{D}_{1z}}-1.96}\right)
\end{equation}
\end{widetext}

\begin{figure}[ptb]
\begin{center}
\includegraphics[clip, width=3.4in ]{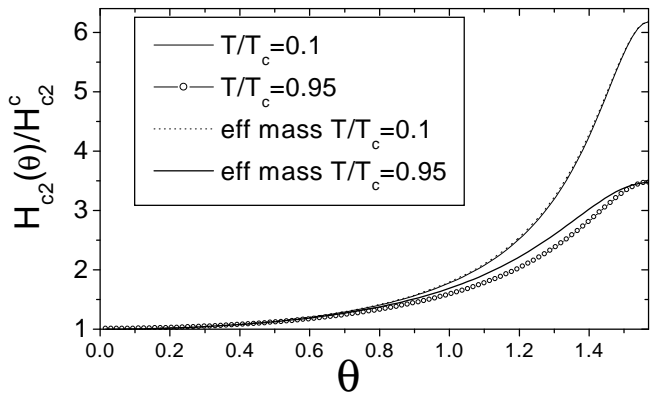}
\end{center}
\caption{Examples of angular dependence of the upper critical
field at low and high temperatures. Fits to the effective-mass
dependence are also shown.} \label{Fig-Hc2angles}
\end{figure}

\begin{figure}[ptb]
\begin{center}
\includegraphics[clip, width=3.4in ]{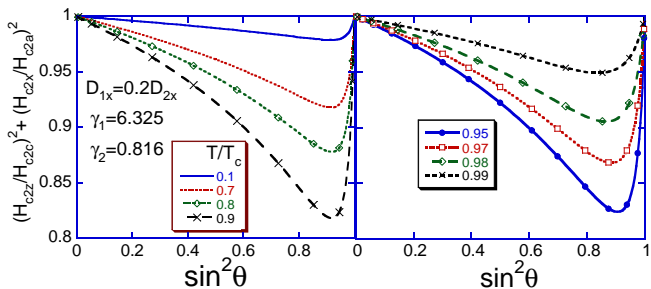}
\end{center}
\caption{Plots of the parameter
$\mathcal{A}(\theta)=(H_{c2,z}(\theta)/H_{c2,c})^{2}+(H_{c2,x}
(\theta)/H_{c2,a})^{2}$ vs $\sin^{2}\theta$ at different
temperatures revealing deviations from the simple effective-mass
law. \emph{Left panel}:  temperatures not very close to $T_c$,
\emph{right panel}: temperature region near $T_c$. }
\label{Fig-effMassDev}
\end{figure}

\begin{figure}[ptb]
\begin{center}
\includegraphics[clip, width=3.4in ]{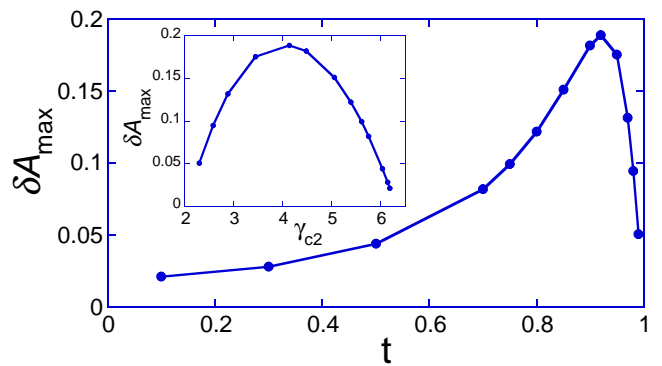}
\end{center}
\caption{Temperature dependence of the parameter
$\delta\mathcal{A}_{\max}=\max_{\theta}\left\vert
1-\mathcal{A}(\theta)\right\vert$ characterizing deviations from
the effective-mass dependence of $H_{c2}$. Inset shows the
dependence of this parameter on the $H_{c2}$-anisotropy.}
\label{Fig-dAmax}
\end{figure}

In the whole angular range we calculated the upper critical field
numerically following the procedure outlined in Sec.\
\ref{Sec:num}. As input parameters we have used the values
$\gamma_{1}=6.325$, $\gamma_{1}=0.816$ which follow from the
electronic band-structure calculations in MgB$_{2}$. We have also
used the relation $\mathcal{D}_{1x}=0.2\mathcal{D}_{2x}$ - the
reason for this choice was discussed in Ref.\ \onlinecite{KG}. The
examples of the calculated angular dependence for $T/T_{c}=0.1$
and $0.95$ are shown in Fig.\ \ref{Fig-Hc2angles} . We also show
fits to a simple effective-mass law, routinely used to describe
angular dependence of $H_{c2}$ in anisotropic superconductors,
$H_{c2}
(\theta)=H_{c2,c}/\sqrt{\cos^{2}\theta+\gamma_{c2}^{-2}\sin^{2}\theta}$.
Due to the contribution from the $\pi$-band, one can see
significant deviations from this law at high temperature. To
enhance these deviations we plot in Fig.\ \ref{Fig-effMassDev} the
angular dependence of the combination $\mathcal{A} (\theta)=\left(
H_{c2,z}(\theta)/H_{c2,c}\right)  ^{2}+\left(  H_{c2,x}
(\theta)/H_{c2,a}\right)  ^{2}$ for several temperatures, (for the
effective-mass law $\mathcal{A}(\theta)=1$ for all $\theta$). We
find that always $\mathcal{A} (\theta)<1$ and the maximum
deviation from unity is achieved around $\theta \thicksim
74^\circ$. At high temperatures one can derive a very simple
formula for $\mathcal{A} (\theta)$ at small angles, $\theta \ll
1$, $\mathcal{A} (\theta)\approx
1-(1/\gamma_{GL}^2-1/\gamma_{c2}^2)\theta^2$. Quantitatively, the
deviations from the effective-mass law can be characterized by the
parameter $\delta\mathcal{A}_{\max}=\max_{\theta}\left\vert
1-\mathcal{A}(\theta)\right\vert$. Fig.\ \ref{Fig-dAmax} shows the
temperature dependence of this parameter. At low temperatures
deviations from the effective-mass law are at the level of several
percents. These deviations progressively grow with the temperature
reaching $\thicksim 19\%$ at $T/T_{c}=0.92$ and then rapidly
decrease when the temperature approaches the narrow GL region near
$T_c$. As the deviations from the effective-mass dependence have
exactly the same origin as the temperature dependence of the
anisotropy, it is interesting to correlate these deviations with
the anisotropy change. Inset in the Fig.\ \ref{Fig-dAmax} shows
plot of the parameter $\delta\mathcal{A}_{\max}$ vs the
$H_{c2}$-anisotropy. One can see that the distortion of the
angular dependence is maximum when the anisotropy is approximately
at the midpoint between the low-temperature and GL limits.
Experimentally, it was found that the angular dependence of
$H_{c2}$ in MgB$_2$ indeed deviates from the effective mass law
\cite{EltsevPhysC02,Welp} and the shape of these deviations
qualitatively agrees with our calculations.

\section{Conclusions}

We have calculated the upper critical field in a dirty two-band
superconductor within the quasiclassical Usadel equations, bearing
in mind the regime of very high anisotropy in the quasi-2D band
relevant for MgB$_{2}$. Following Ref.\ \onlinecite{Mazin02}, we
have assumed that the interband scattering is negligible even in
the dirty limit in both bands. Most of MgB$_{2}$ samples are in
dirty limit, except for single crystals, where the dirty limit
conditions are fulfilled in the $\pi$-band but not fulfilled in
the $\sigma$-band.\cite{YelandPRL02} Still, as argued in Ref.\
\onlinecite{KG}, our results should be qualitatively applicable to
MgB$_{2}$ single crystals, if one considers the coherence length
$\xi_{1}$ as a phenomenological parameter instead of expressing it
via the diffusion constant $\mathcal{D}_{1}$.

We have considered separately the cases when the field is parallel
and perpendicular to the basal plane. We have found that at low
temperatures both critical fields are mainly determined by the
strong band and only weakly deviate from the universal Maki - de
Gennes result. The low temperature anisotropy is mainly determined
by the anisotropy of diffusion constants in a
quasi-two-dimensional band. However, the anisotropy is suppressed
at high temperatures. The reason is that there are two important
parameters, anisotropy of pairing interaction and of diffusion
constants, which enter the expression for the parallel $H_{c2}$ in
a different way at high and low temperatures. This property can be
expressed as the anisotropy of coherence length $\xi_{ab}/\xi_{c}$
which decreases with increasing temperature. This effect is in
accordance with the experimental data in MgB$_{2}$. Note that the
anisotropy of the penetration depth $\lambda_{c}/\lambda_{ab}$
increases with increasing temperature,\cite{Kogan_lam,Golub_lam}
which is another manifestation of the two-band model.

We have also studied quantitatively the dependence of case of
$H_{c2}$ on the angle between the \textit{ab}-plane and the
magnetic field direction. Approximate relations for $H_{c2}$
dependence on titled angle are derived for small and large angles.
In the whole angular range numerical calculations are performed.
The results demonstrate the deviation from the effective-mass
dependence. This means the breakdown of anisotropic GL theory.
Further, we have shown that the temperature range of applicability
of the GL theory is extremely narrow in the considered two-band
case.

Another issue is strong coupling corrections to $H_{c2}$. In this
paper the weak coupling approach was used. On the other hand, it
is known from work on isotropic superconductors \cite{SD} that
strong coupling corrections renormalize the
 absolute value of $H_{c2}$ by the factor $(1+\lambda)^{\alpha}$,
where $\lambda$ is the coupling constant and $\alpha\simeq2$.
Since electron-phonon coupling in MgB$_{2}$ is relatively strong
(according to Ref.\ \onlinecite{Golub}, $\lambda_{11}\simeq1$),
these corrections are important for calculation of absolute values
of $H_{c2}$. However, we do not expect qualitative changes in the
temperature and angle dependencies of the anisotropy ratio
calculated in the present paper. Extension of our results to the
strong coupling Eliashberg regime is an interesting subject for
future work.

We acknowledge valuable discussions with A.\ Brinkman, O.\ V.\
Dolgov, I.\ I.\ Mazin, U.\ Welp, A.\ Rydh, M.\ Iavarone, and G.\
Karapetrov. In Argonne this work was supported by the U.S. DOE,
Office of Science, under contract \# W-31-109-ENG-38.

\appendix

\section{\label{App:Local} Local approximation for the $\pi$-band}

Let us consider equation for the $F$-function in the weak $\pi$-band
\begin{equation}
\omega F_{2}+\frac{\mathcal{D}_{2z}}{2}\left(  \frac{2\pi
Hx}{\Phi_{0} }\right)
^{2}F_{2}-\frac{\mathcal{D}_{2x}}{2}\nabla_{x}^{2}F_{2}=\Delta_{2}.
\label{F-eq-pi}
\end{equation}
The typical scale of $\Delta_{2}(x)$ variation is imposed by the
strong $\sigma$-band. This scale is given by
$x_{1}=\sqrt{\gamma_{1}\Phi_{0}/2\pi H}$ and, due to inequality
$\gamma_{1}\gg\gamma_{2}$, it is much larger than the length scale
$x_{2}=\sqrt{\gamma_{2}\Phi_{0}/2\pi H}$ of the oscillator
operator in the left side of the Eq.\ (\ref{F-eq-pi}). For
relevant $\omega$'s the typical length scale of $F_{2}$ variation
is much larger than $x_{2}$. This allows us to neglect the
gradient term in Eq.\ (\ref{F-eq-pi}). This approximation is
equivalent to the approximation for the matrix elements used in
Eqs.\ (\ref{MatrixElemApprox}). Then the $\pi$-band $F$-function
is given by
\begin{equation}
F_{2}=\frac{\Delta_{2}}{\omega+\frac{\mathcal{D}_{2z}}{2}\left(  \frac{2\pi
Hx}{\Phi_{0}}\right)  ^{2}}. \label{F2Local}
\end{equation}
Substituting this expression into the second self consistency equation, we
represent it in the form
\begin{equation}
-W_{21}\Delta_{1}+W_{2}\Delta_{2}=\left(  \ln\frac{1}{t}-g\left[
\frac{\mathcal{D}_{2z}}{4\pi T}\left(  \frac{2\pi Hx}{\Phi_{0}}\right)
^{2}\right]  \right)  \Delta_{2}. \label{SelfConsLocal}
\end{equation}
It has to be solved together with equations for $F_{1}$ and the first self
consistency equation. Using expansion with respect to eigenfunctions of the
$\sigma$-band $\Psi_{n}(x)$, this equation reduces to the form of linear
equation (\ref{SelfConsReducedB}), in which the matrix $U_{i,j}$ is given by
the matrix elements
\[
U_{i,j}=-\int dx\Psi_{2i}(x)\Psi_{2j}(x)g\left[  \frac{\mathcal{D}_{2z}}{4\pi
T}\left(  \frac{2\pi Hx}{\Phi_{0}}\right)  ^{2}\right]  .
\]
Introducing the dimensionless oscillator wave functions
$\psi_{n}(\tilde{x})$,
$\Psi_{n}(x)=\psi_{n}(x/x_{1})/\sqrt{x_{1}}$, we present these
matrix elements in the dimensionless form
\[
U_{i,j}(t_{2})=\int_{-\infty}^{\infty}d\tilde{x}\psi_{2i}(\tilde{x})\psi
_{2j}(\tilde{x})g\left[  \tilde{x}^{2}/t_{2}\right]  ,
\]
where, again $1/t_{2}\equiv\left(
\mathcal{D}_{2z}/\mathcal{D}_{1z}\right) \left(  H/tH_{1}\right)
$. In particular, $\psi_{0}(\tilde{x})=\pi^{-1/4}
\exp(-\tilde{x}^{2}/2)$. In the "high-field in $\pi$-band" regime,
$t_{2}\ll 1$, one can use asymptotics $g\left[
\tilde{x}^{2}/t_{2}\right]  \approx 2\ln(\tilde{x})-\ln
t_{2}+\gamma_{E}+2\ln2$ and obtain
\begin{align*}
U_{i,j}(t_{2})  &  =f_{i,j}+(\ln t_{2}-\gamma_{E}-2\ln2)\delta_{n,m}\\
f_{i,j}  &
=-4\int_{0}^{\infty}d\tilde{x}\psi_{2i}(\tilde{x})\psi_{2j}
(\tilde{x})\ln(\tilde{x}).
\end{align*}
In particular,
\[
f_{0,0}=-\frac{4}{\sqrt{\pi}}\int d\tilde{x}\exp(-\tilde{x}^{2})\ln(\tilde
{x})=\gamma_{E}+2\ln2\approx1.9635
\]


\begin{thebibliography}{99}
\bibitem{Parameters}The parameter $S_{12}$ is related with the parameter
$\eta$ from Ref.\ \onlinecite{Dahm} as $S_{12}=\eta/(1-\eta)$.
\bibitem{GurevichError}This limit corresponds to Eq.\ (35) of
Ref.\ \onlinecite{Gurevich}. However this formula is written with a typo:
there should be a minus sign in front of $\lambda_{-}$ inside the exponential factor.
\end{thebibliography}
\end{document}